\documentclass[showpacs,preprintnumbers,amsmath,amssymb,prl,twocolumn]{revtex4}

\usepackage{amsmath}
\usepackage{amssymb}
\usepackage{color}
\usepackage{graphicx}
\usepackage{epstopdf}
\usepackage[english]{babel}

\newcommand{\ket}[1]{\left\vert#1\right\rangle}

\begin{document}
\title{Experimental Adaptive Bayesian Tomography}
\author{K.S.Kravtsov$^1$}\author{S.S.Straupe$^2$}\email{straups@yandex.ru} \author{I.V.Radchenko$^1$, G.I.Struchalin$^2$, N.M.T.Houlsby$^3$, S.P.Kulik$^2$}
\affiliation{$^1$ A.M.Prokhorov General Physics Institute RAS, Moscow, Russia \\
$^2$Faculty of Physics, M.V.Lomonosov Moscow State University, Moscow, Russia\\
$^3$ Computational and Biological Learning Lab, Department of Engineering, University of Cambridge, Cambridge, United Kingdom}

\date{\today}
\begin{abstract}

We report an experimental realization of an adaptive quantum state tomography protocol. Our method takes advantage of a Bayesian approach to statistical inference and is naturally tailored for adaptive strategies. For pure states we observe close to $1/N$ scaling of infidelity with overall number of registered events, while best non-adaptive protocols allow for $1/\sqrt{N}$ scaling only. Experiments are performed for polarization qubits, but the approach is readily adapted to any dimension.

\end{abstract}
%\pacs{03.67.Bg, 03.67.Mn, 42.65.Lm}
\maketitle

The main goal of quantum state tomography is to provide an estimate $\hat\rho$ for an unknown quantum state $\rho$ based on the data collected in a series of measurements \cite{Paris_04}. The estimator is supposed to be close to the real state in some reasonable sense, therefore various notions of statistical distance between quantum states are used \cite{Caves_PRL94, Bengtsson_06}. One of the possible measures of statistical distance is \emph{infidelity} \footnote{Sometimes, eg. in \cite{Kulik_PRA11} this quantity is called \emph{fidelity loss.}}, defined as $1-F(\rho,\hat{\rho})=1-\mathrm{Tr\left(\sqrt{\sqrt{\rho}\hat\rho\sqrt{\rho}}\right)}^2$. The ultimate goal of any tomographic protocol is to minimize this distance for a fixed overall number of measurements made $N$. Usually a protocol makes use of some fixed number of measurement settings predetermined before the actual experiment. For such a protocol the infidelity scales as $1-F\sim N^{-1/2}$ for the most interesting for applications set of almost-pure states. One can more or less significantly alter the pre-factor by a clever choice of measurements \cite{Rehacek_PRA04, Langford_PRA08, Kulik_PRA11}, but the scaling is unaffected. A natural question is whether it is possible to beat this limit? The answer turns out to be positive if one allows for adaptivity - the measurement performed at some step of the protocol should be determined in dependence of the data obtained in the previous ones \cite{Freyberger_PRA00, Wunderlich_PRA02}.

Here we report an experimental approach to adaptive quantum state tomography based on a recently proposed adaptive Bayesian estimation algorithm \cite{Houlsby_PRA12}. We achieve almost $1/N$ scaling of infidelity for pure states of polarization qubits and demonstrate a clear advantage over best symmetric non-adaptive protocols. Our approach is completely different and more general than that of another recent experimental realization \cite{Takeuchi_PRL12}, where adaptive measurements were used to estimate a single unknown parameter of a quantum state.

\paragraph{Bayesian tomography.}
Let us start with describing a general framework for quantum state estimation and, in particular, the Bayesian approach. A tomographic protocol is a set of positive operator valued measures (POVM's) $\mathcal{M}=\left\{ \mathbb{M}_{\alpha}\right\}$ with index $\alpha$ numbering the different configurations of the experimental apparatus. In a given configuration, the probabilistic outcome of each measurement $\gamma$ being observed is determined according to the Born's rule:
\begin{equation}\label{Born}
\mathbb{P}(\gamma|\rho,\alpha)=\mathrm{Tr}[M_{\alpha\gamma}\rho],
\end{equation}
where $M_{\alpha\gamma}$ are POVM elements, obeying $\sum_{\gamma=0}^{\Gamma-1} {M_{\alpha\gamma}}=I$, and $\rho$ is the density matrix of the state to be determined. The set $\mathcal{D}$ of all outcomes observed in an experiment form the data set used to estimate density matrix elements. The Bayesian approach to statistical inference dictates the following rules:
\begin{itemize}
    \item{a \emph{prior} distribution over the space of density matrices $p(\rho)$ is specified;}
    \item{the collected data are used to obtain the posterior distribution $p(\rho|\mathcal{D})\propto\mathcal{L}(\rho;\mathcal{D})p(\rho)$, $\mathcal{L}(\rho;\mathcal{D})$ is the \emph{likelihood} function, and it contains our statistical model that encodes probabilistic mapping from the state to the observed data;}
    \item{quantities of interest are estimated using expected values under the posterior distribution: for example, we may obtain the \emph{Bayesian mean estimate} of the state as $\hat{\rho}=\mathbb{E}_{p(\rho|\mathcal{D})}\left[\rho\right]$. Variance, infidelity or any other statistical quantity of interest may be obtained in the same way.}
\end{itemize}

The Bayesian approach has many advantages over a more standard for quantum information community maximum-likelihood estimation (MLE) \cite{Hradil_PRA97}. It offers, in a natural way, a distribution over the space of density matrices, which provides the most complete description of our knowledge about the quantum state, inferred from data $\mathcal{D}$ \cite{Blume-Kohout_NJP10}. Even more importantly for us, it is a natural framework for construction of adaptive estimation protocols. Indeed, the posterior distribution may be updated as soon as one observes some data, in the extreme case~-- after each measurement, and the new knowledge about the state may be used to select the next measurement setting $\alpha$ in a most optimal way. Choosing the criterion for ``optimality" is a task of \emph{optimal experiment design} (OED) and may be solved in various ways. In Bayesian framework a natural strategy is to choose a measurement, maximally reducing the entropy of the posterior~-- it means, that our knowledge about the state, obtained after such measurement is maximized \cite{Houlsby_PRA12}. This may be formulated as choosing a measurement configuration $\alpha$ as a solution to the following optimization procedure:
\begin{equation}\label{Objective}
    \alpha=\mathrm{arg}\max\limits_{\alpha\in\mathcal{A}} \left\{ \mathbb{H}\left[p(\rho|\mathcal{D})\right] - \mathbb{E}_{p(\gamma|\alpha,\mathcal{D})} \mathbb{H}\left[ p(\rho|\gamma,\alpha,\mathcal{D}) \right] \right\}.
\end{equation}
Note, that because we do not know which outcome $\gamma$ will be observed we use the \emph{expected} information gain (under the posterior) as our objective. Before describing how to work with \eqref{Objective} in practice we detail the components required for our Bayesian model.

\paragraph{The likelihood function.}
The likelihood function is equal to the probability of the observed data, given a particular state, i.e. $\mathcal{L}(\rho;\mathcal{D})=\mathbb{P}(\gamma|\rho,\alpha)$. In the simplest setting, i.e. in the absence of any experimental noise, the likelihood function is given directly by Born's rule \eqref{Born}. In practice experimental noise also needs to be modelled in the likelihood function, we present these extensions to the simple model later in the paper.

\paragraph{The prior.}
As the Bayesian framework implies finding a probability distribution instead of a single point estimate, the analysis should also take into account the particular geometry of the space, i.e. the geometry of single qubit density matrix space. In general, the geometry of space is defined by its metric, which provides a notion of distance. In the case of density matrices the natural choice of metric is Bures distance, defined as $d_B^2(\rho_1,\rho_2) = 2 - 2\sqrt{F(\rho_1,\rho_2)}$. Locally it coincides with the concept of Fubini-Study (quantum angle) measure and the Hilbert-Schmidt distance (see Ref. \cite{Bengtsson_06}). As follows from its definition, for close-by states the Bures distance is a square root of infidelity.

It can be shown that the curvature of Bures metric space for single qubit states is constant \cite{Dittmann_JGP99}, thus it can be isometric to a hypersphere in a 4-d space. A simple isometric mapping exist between the hemisphere of radius 1/2 and the Stokes parameters: \begin{equation}\begin{array}{l} x_1 = \frac12 S_1;\quad x_2 = \frac12 S_2;\quad x_3 =\frac12 S_3;\\ x_4 =\frac12 \sqrt{1-S_1^2- S_2^2 - S_3^2}.\end{array}\label{3-sphere}\end{equation} Thus a uniform distribution in the Bures distance sense is a projection of a uniformly populated 3-sphere to the space of Stokes parameters. This projection apparently gives a lower density in its center and a higher near its surface.

The first step in any Bayesian estimation procedure is to choose a prior distribution. In the ideal situation a prior should give absolutely no information about the system, thus a typical choice for prior is a Haar measure in the space in question. In our case we use Haar measure in Bures metric space. However, to stick with conventional (and very convenient) parameterization with three Stokes parameters we use the above mentioned isometry from a 3-sphere. Samples obtained by this procedure concentrate more along the surface of the Stokes parameters ball, which is natural in the sense that the distance and, thus, {\em fidelity} between the samples remains uniform over the whole space. This would not be the case if the ball was populated uniformly, as fidelity is not directly connected with separations in the space of Stokes parameters.

We should also mention here another strategy extensively used in literature for prior generation \cite{Blume-Kohout_NJP10}. According to it a density matrix for a single-qubit is treated as a result of tracing out other qubits from a higher-dimensional pure state. Starting from different dimensions one gets different prior distributions, however with the increase of dimensionality the distribution tends to concentrate around the completely mixed state giving less and less chances to pure states. This unnatural behavior prevents us from using it.

\begin{figure}[h!]
\includegraphics[width=0.32\columnwidth]{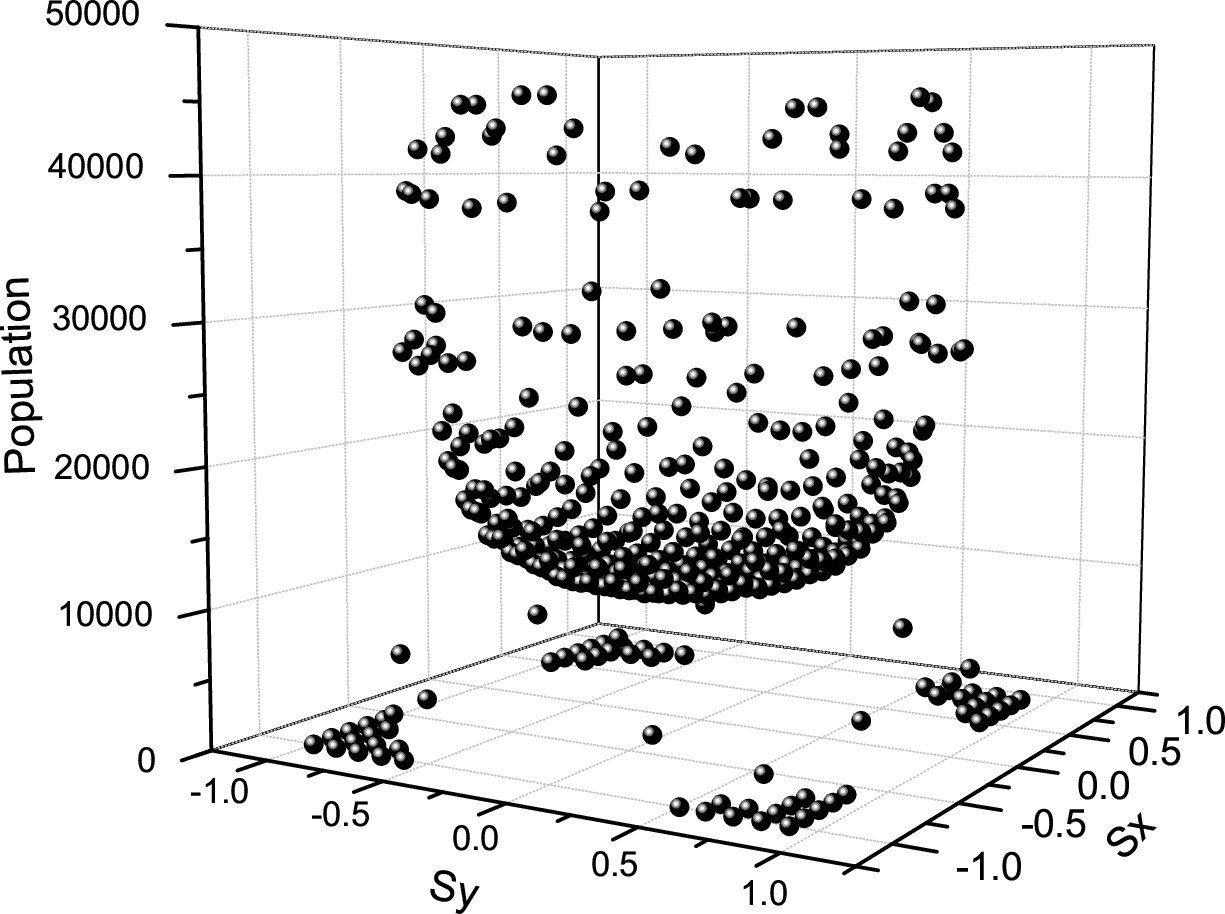}
\includegraphics[width=0.33\columnwidth]{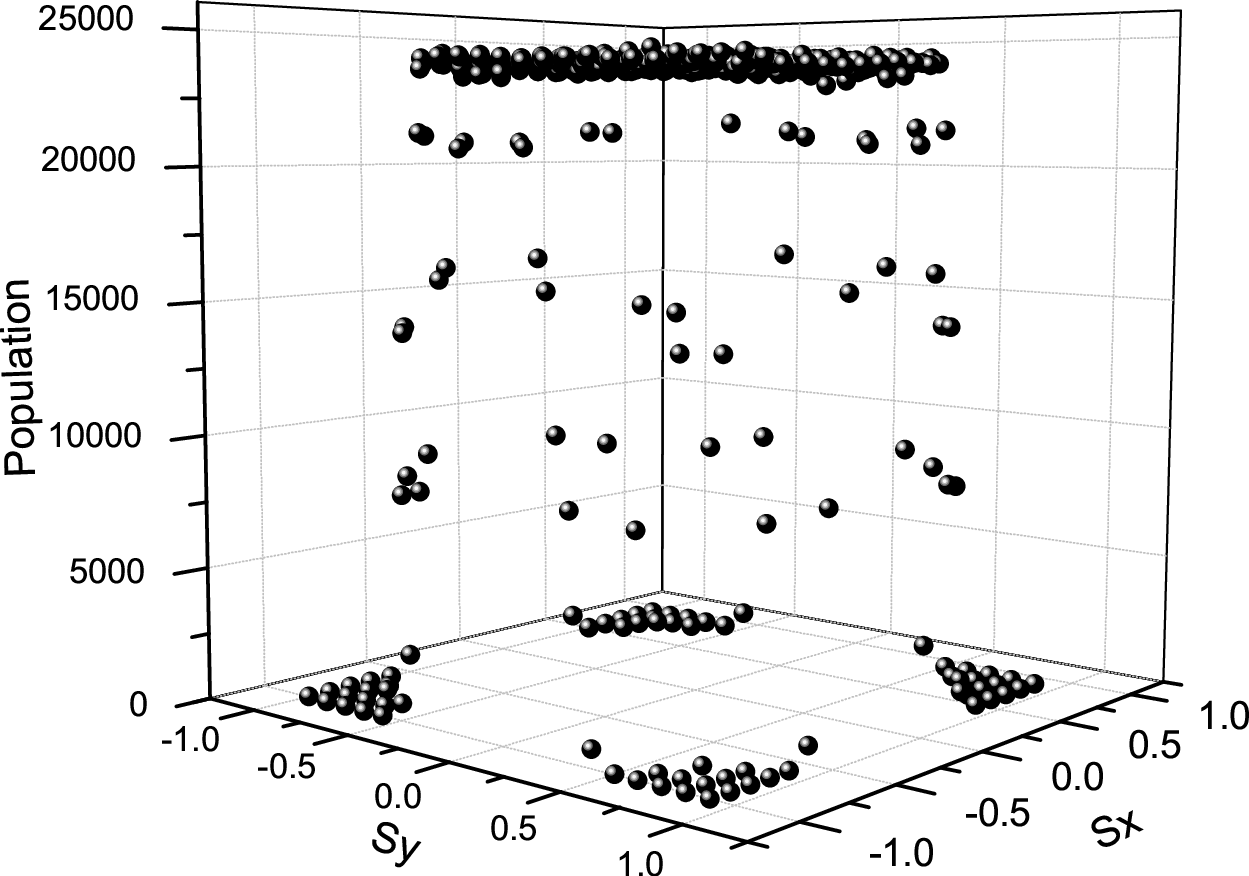}
\includegraphics[width=0.30\columnwidth]{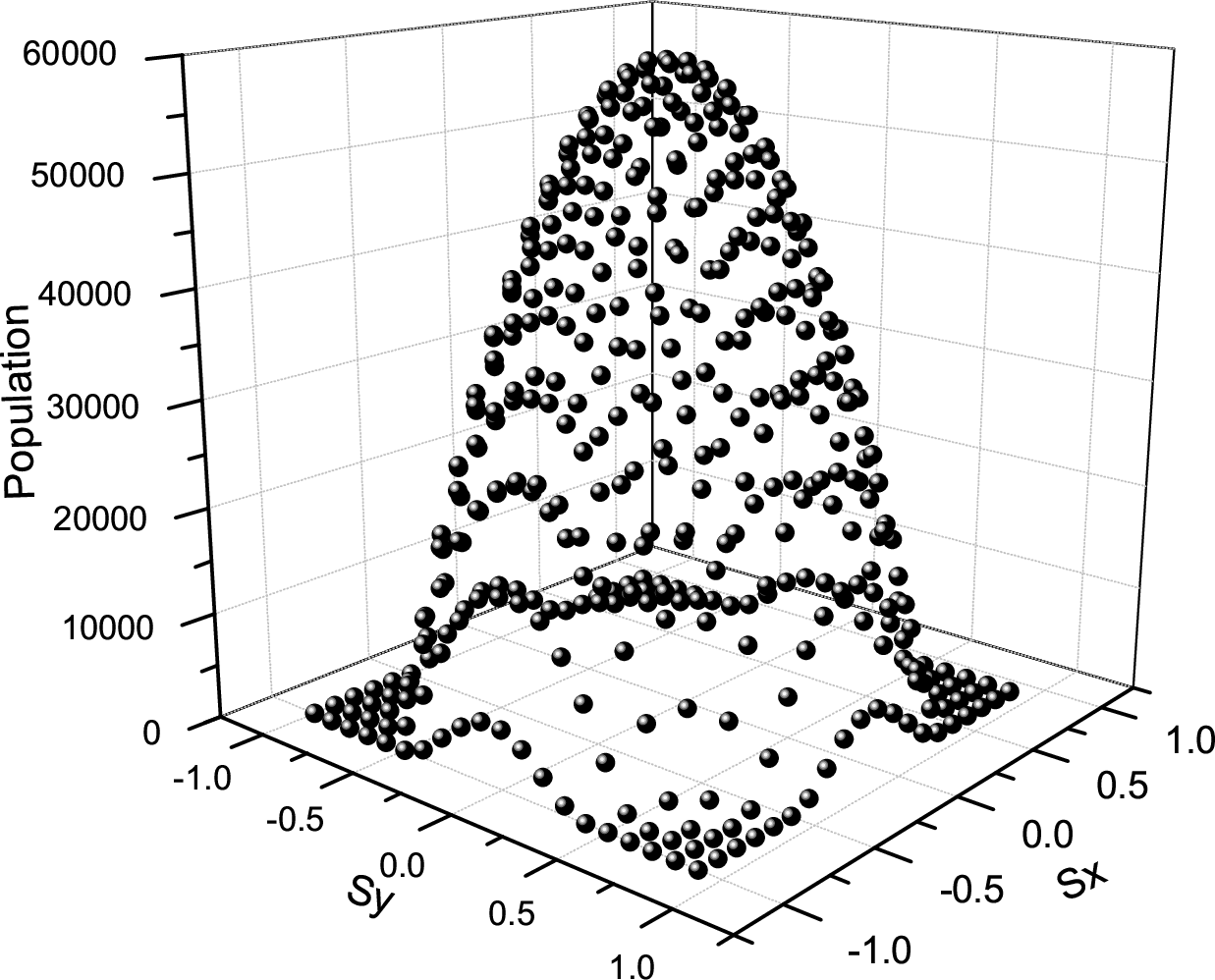}
a)\hspace{0.3\columnwidth}b)\hspace{0.3\columnwidth}c)
{\caption{Population of samples in the space of Stokes parameters vs. $S_x$ and $S_y$ condition to $|S_z|<0.05.$ Samples are derived from (a) Haar measure in Bures metric space; two qubit (b) and qubit-qutrit (c) pure states traced by the second particle. The Haar measure gives flat infidelity distribution between all samples, while the latter two favor mixed states giving poorer fidelity to pure states.}\label{Densities}}
\end{figure}

To illustrate the above, in Fig.~\ref{Densities} we show density of samples in a flat disc cut from the center of the Stokes parameters ball for the Haar measure (a), and for the results of tracing down from a two-qubit (b) and a qubit-qutrit (c) pure states. The first distribution has more density towards the circumference following the behavior of fidelity in the Stokes parameter space. Other distributions give more and more favor to mixed states, unveiling their inadequateness for our purpose.

\paragraph{Approximate Inference.}
One of the principle reasons that Bayesian methods enjoy less popularity in quantum tomography than MLE is the fact that posterior normalization requires computing an (in general high-dimensional) integral of likelihood function, which is computationally hard. Usually, when faced with intractable Bayesian inference, the posterior is approximated via sampling \cite{blume2010}, or by approximating the posterior with a simpler distribution, such as a Gaussian \cite{audenaert2009}.

This computational difficulty is further compounded when one is performing adaptive quantum tomography, one must keep track of the current posterior after making $n$ measurements, in order to calculate the optimal $n+1$'st measurement. To perform inference based on all the observed data is at best an $\mathcal{O}(n)$ operation, which becomes increasingly problematic as the experiment progresses. Fortunately fast algorithms for solving online Bayesian inference problems exist; they update the posterior after inclusion of each new datapoint, without re-visiting all previous data. We briefly review the core idea behind this approach, and refer the reader to \cite{doucet2001} for details.

The algorithm is a variant of sequential importance sampling algorithm (SIS) with resampling. The idea is to construct a particle filter, approximating the posterior with a set of weighted samples, i.e. $p(\rho|\mathcal{D}_n)\approx\sum_{s=1}^{S} w_s^{(n)}\delta(\rho-\rho_s)$. After each observation one updates the weights $w_s^{(n)}$; this can be done incrementally, using the current set of particles and weights, and the likelihood corresponding to the new observation. The algorithm has $O(1)$ cost for integration of the likelihood, which means that it can be applied on-line at every step of the adaptive protocol, irrespective of the current amount of data collected.

A common problem with weighted particle filters is that after collection of sufficient data, a small collection of particles get almost all of the weight, this means that the \emph{effective sample size} reduces and the quality of the approximation to the posterior becomes poor. This problem is avoided by monitoring the effective sample size and re-sampling the particles when it falls too low. The resampling procedure uses two phases, firstly the particles are re-drawn from the set of current particles in proportion to their weight (and the weights are equalized), then they are ``spread back out'' using the Metropolis Hasting's algorithm. It is important to note that implementation of Metropolis-Hastings algorithm is convenient to perform on a 3-sphere surface defined by~(\ref{3-sphere}), which automatically ensures correct step sizes and avoids unnecessary boundary conditions on the surface of the Bloch ball, which, in general, may bias the posterior. Using the particles, whenever we require an expectation under the posterior, e.g. when computing the mean fidelity, or when computing the next state for adaptive tomography, one can simply replace the complex integrals with simple weighted Monte Carlo estimates.

\paragraph{Efficient Adaptive Tomography.}
We now return to computation of the objective function for adaptive tomography (\ref{Objective}), although this objective is theoretically attractive, even with the sampling estimate of the posterior two major computational difficulties are encountered. Firstly, we must compute entropies of (in general) high-dimensional quantum states; it is notoriously hard to compute entropies directly from samples from the distribution \cite{panzeri2007}. Secondly, one requires the posterior distribution $p(\rho|\gamma,\alpha,\mathcal{D})$ for all possible next measurements $\alpha$, and all their possible outcomes $\gamma$; this would require performing an SIS update for all these possible scenarios.
This computational burden is unavoidable if one uses a loss function other that the log loss, which leads to the Shannon's entropy objective function, which means that optimal designs can only be computed for very short experiments \cite{hannemann2002}. Therefore, it is highly beneficial to work with the following equivalent formulation:
\begin{equation}\label{Objective2}
    \alpha=\mathrm{arg}\max\limits_{\alpha\in\mathcal{A}} \left\{ \mathbb{H}\left[p(\gamma|\alpha,\mathcal{D})\right] - \mathbb{E}_{p(\rho|\mathcal{D})} \mathbb{H}\left[ p(\gamma|\alpha,\rho) \right] \right\}.
\end{equation}

In \eqref{Objective2} only predictive entropies are required, which is much easier because output space is typically much lower dimensional than state space, and only the \emph{current} posterior is needed $p(\rho|\mathcal{D})$.

\paragraph{Simulations.}
We performed simulated experiments to empirically evaluate the performance of Bayesian adaptive tomography. For our performance metric we use the mean infidelity as measured against the true state, $\bar{\rho}$: $1-\hat{F}(\rho,\bar{\rho}) = \mathbb{E}_{p(\rho|\mathcal{D}_n)}\left[ 1-F(\rho,\bar{\rho}) \right]$. Note that Bayesian mean $1-\hat{F}(\rho,\bar{\rho})$ is a ``fairer'' score than the fidelity of a point estimate, e.g. the posterior mean (i.e.  $1-F(\mathbb{E}_{p(\rho|\mathcal{D}_n)}[\rho],\bar{\rho})$). The fidelity of the posterior mean does not take into account the uncertainty captured by the posterior. The posterior mean could be correct, for example, if the state is pure and the posterior has become ``flattened'' against the surface of the Bloch sphere; the variance, however, could be very high - the system may have little knowledge of the polar and azimuth angles (in the Bloch sphere) of the true state. The Bayesian estimator rewards posterior distributions that are centered in the correct location \emph{and} have low variance.

\begin{figure}[h!]
\includegraphics[width=\columnwidth]{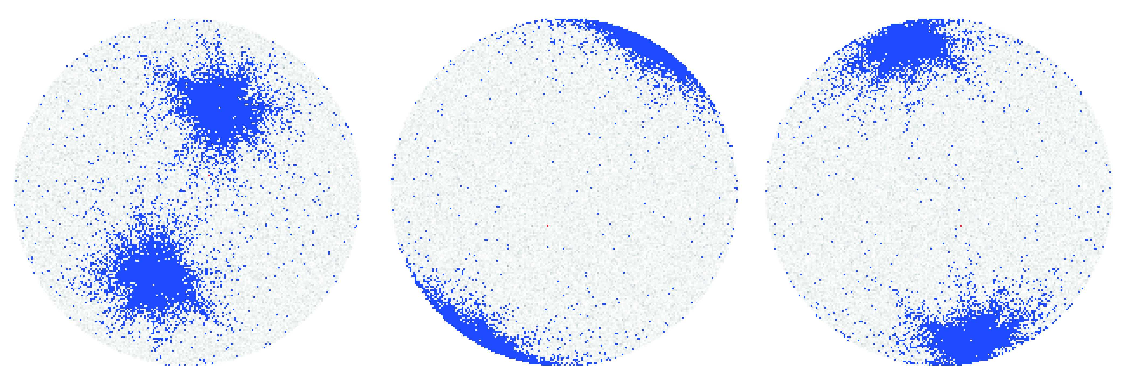}
{\caption{Illustration of an adaptive choice of measurements at reconstruction of a pure state. The Bloch sphere is given in three orthogonal projections with each measurement basis marked as a dark point. As expected, measurement bases tend to concentrate around the reconstructed state or the symmetric state on the other side of the sphere. }\label{Measure}}
\end{figure}

To achieve statistically significant results we perform multiple runs within each simulation. For each run we generate a random pure state which we use as the true state $\bar{\rho}$. We average over 20 runs, each with a different true state. All measurements performed in a single run are shown in Fig.~\ref{Measure}. After the system collects some initial information about the measured state it tends to choose measurements aligned with its current estimate of the true state (although not exclusively), thus taking advantage of the adaptive approach. Interestingly, the optimal first three measurements chosen correspond precisely to a set of Mutually Unbiased Bases (MUBs), but diverge from these MUBs on the 4th. We compare the results to measurements chosen randomly and uniformly, and selecting randomly from a set of MUBs, which have been shown to be the optimal (in terms of information gain) non-adaptive measurement set one can choose prior to the experiment \cite{wootters1989}. We also fit a power law, of the form $1-\hat{F}\propto N^a$ to the data, in order to compare the convergence rates of the different methods. Note that the performance of the random and adaptive schemes is independent of the angle of the true state (the algorithms remain the same given any rotation of the Stokes coordinate), however, the fixed MUBs are not. Drawing intuition from the fact that the adaptive scheme selects mostly measurements that align with the true state (hence ``squashing'' the posterior against the Bloch sphere), we find that the ``best case'' for MUB tomography is when the state is aligned with one of the MUB measurements~
%%%%%%%%%%%%%%%%%%%%%
%Gleb
\cite{Steinberg}
%%%%%%%%%%%%%%%%%%%%%
, and the worst case is when the state is equally biased to all measurements i.e. $\{S_x=S_y=S_z=1/\sqrt{3}\}$.

\begin{figure}[h!]
\includegraphics[width=\columnwidth]{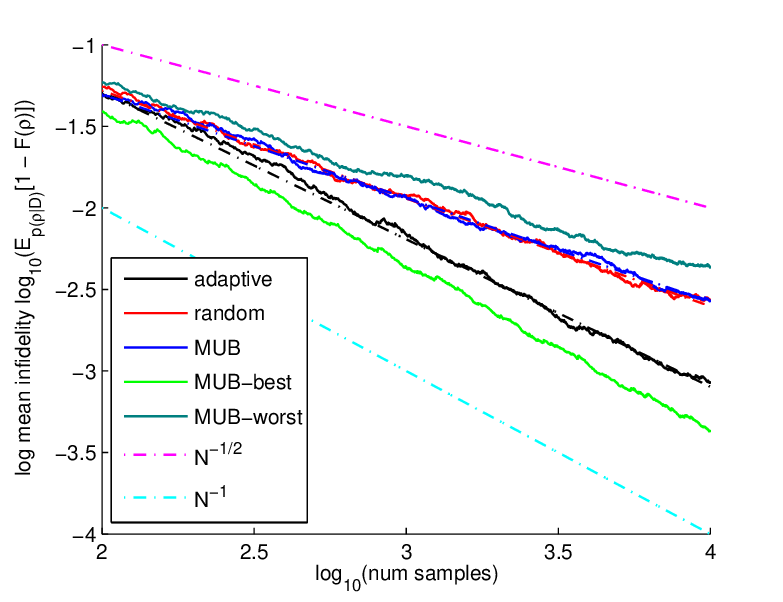}
{\caption{Simulated tomography using three measurement selection methods: randomly sampled (red continuous line), MUBs (blue $\times$) and fully adaptive Bayesian tomography (black $\circ$). For these methods, the true state is random and pure, the results presented here are the average of 20 independent runs. Overlaid dashed lines indicate the power law fit. Functions $1-F=N^{-1/2}$ (magenta, dash-dotted) and $1-F=N^{-1}$ (cyan, dashed) are shown for comparison.
To account for state-dependence of MUB tomography, we also present its performance for the ``worst'' and ``best'' (see text) true states (dark green $+$, light green $\bullet$, respectively).
}\label{Simulations}}
\end{figure}

The results are presented in Fig.~\ref{Simulations}. Firstly, note that random tomography yields a $1/\sqrt{N}$ rate as expected, $a=-0.448\pm0.183$. However, adaptive tomography performs close to the at $1/N$ level on average $a=-0.915\pm0.101$. In its most favorable scenario, MUBs also perform close to the $1/N$ rate (with a small multiplicative constant improvement over the adaptive scheme). In practice, because the optimal MUBs depend on the state, these are not known a priori. In the worst case MUBs achieve $1/\sqrt{N}$ convergence; but we also observe that on average the rate is closer to $1/\sqrt{N}$ than $1/N$, $a=-0.593\pm0.134$.

\begin{figure}[h!]
\includegraphics[width=\columnwidth]{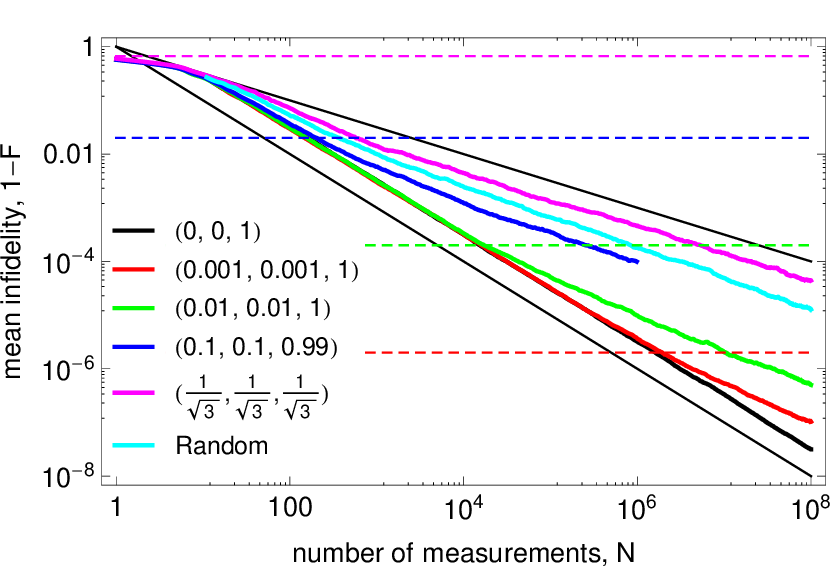}
{\caption{Convergence scaling transformation for different relative orientation of the true state and the MUBs. Thick lines~-- mean infidelity $1-\hat F(\rho,\bar\rho)$ of the current Bayesian mean estimate with the true state. Each curve is a result of averaging over 100 simulation runs. The true state $\bar\rho$ is fixed and has Stokes parameters $S_x=0, S_y=0, S_z=1$. Legend shows the values of the Stokes parameters for the MUB vector closest to the true state $\sigma_{near}$. Infidelity between the true state and the nearest vector $1-F(\bar\rho, \sigma_{near})$ is depicted by horizontal dashed lines of appropriate colors. Thin black lines are functions $1/\sqrt{N}$ and $1/N$ shown for comparison. Transformation of infidelity scaling $1/N \to 1/\sqrt{N}$ takes place exactly when infidelity $1-\hat F(\bar\rho, \rho)$ equals $1-F(\bar\rho, \sigma_{near})$.} \label{ConvergenceMUBs}}
\end{figure}

%%%%%%%%%%%%%%%%%%%%%
%Gleb: новый раздел
It is interesting to consider the convergence of MUB-based protocols in the intermediate case between ``worst'' and ``best'' MUBs orientation and estimate the sensitivity to basis misalignment. Fig.~\ref{ConvergenceMUBs} shows the results of simulations for different relative orientation of the true state $\bar\rho$ and the measurement basis. In these simulations the true state was fixed and orientation of the measurement basis with respect to this state was varied. Let us denote the closest to the true state vector of MUB as $\sigma_{near}$. $1/N$ convergence law was observed for all MUB orientations (except the ``worst'' case) until the number of measurements performed reached a certain threshold $N_{thrs}$, for larger number of measurements the scaling law for infidelity changed to $1/\sqrt{N}$. The actual value of the threshold depends on the basis misorientation, quantified by the infidelity between the true state and the closest state of the MUB $1-F(\bar\rho, \sigma_{near})$.

With only two characteristic quantities, namely the mean infidelity with the true state $1-\hat F(\rho, \bar\rho)$ and the infidelity of the true state and the nearest MUB vector $1-F(\bar\rho, \sigma_{near})$ at hand, one can conjecture that the scaling transformation occurs when these values coincide. The underlying reasoning is quite simple: if $1-\hat F(\rho, \bar\rho)$ is sufficiently large (greater than $1-F(\bar\rho, \sigma_{near})$) then MUBs are still aligned with true state with accuracy exceeding our current knowledge of true state~-- it is the ``best'' case for MUB measurements. Otherwise, if $1-\hat F(\rho, \bar\rho)$ is small, the misalignment becomes noticeable since the nearest vector $\sigma_{near}$ lies outside the region where the true state is likely situated. Simulation results shown in Fig.~\ref{ConvergenceMUBs} confirm this hypothesis.

%%%%%%%%%%%%%%%%%%%%%

\paragraph{Measurement blocks.}
One can save computational and experimental time by taking blocks of measurements using the same configuration for $k$ consecutive measurements. As the posterior collapses towards the true state, the optimal measurement changes less frequently - a direction pointing towards the true state becomes increasingly preferable. Therefore, as the experiment progresses the block size $k$ can be allowed to grow without detrimenting the quality of the experimental procedure. We use a heuristic block-size schedule which increases the block size at a $\mathcal{O}(n)$ rate, where $n$ is the number of measurements seen so far. In particular we use $k=\text{max}(\lfloor n/100 \rfloor, 1)$; in this case the achieved infidelity scales linearly with elapsed time. Simulations show that using this schedule does not make any statistically significant difference to the convergence rate, or even to the absolute fidelity achieved at any time.
%\begin{figure}[h!]
%\includegraphics[width=\columnwidth]{Simulation1}
%{\caption{Description.}\label{Simulation1}}
%\end{figure}

%\begin{figure}[h!]
%\includegraphics[width=0.8\columnwidth]{comparison}
%{\caption{\textcolor{blue}{Simulation results for various types of non-adaptive and adaptive protocols. Stokes parameters measurements (measurements in a Pauli basis) (red line) and completely random measurements (green line) exhibit $N^{-1/2}$ behavior, while adaptive measurements (blue line) clearly demonstrate $N^{-1}$ scaling. Straight solid lines are functions $1-F=N^{-1/2}$ (black) and $1-F=N^{-1}$ (purple), shown for comparison.} }\label{Simulation2}}
%\end{figure}

\paragraph{Experimental imperfections.}
In practice quantum tomography is inevitably subject to experimental noise. This noise is not modeled in the likelihood function given by Born's rule \eqref{Born}, and therefore, its presence may bias the results of inference, reducing both the fidelity of the inferred state, and the optimality of the adaptive experimental design. For our set-up we have identified two major additional sources of experimental noise. Firstly, the presence of detector dark counts with detector-specific rates. Secondly, attenuation in both channels due to detector inefficiency and losses/reflections at the optical elements. If the attenuation was equal in both channels, then the inference would be unaffected; however, unequal attenuation will bias the posterior.

A popular approach to modeling the additional uncertainty in the state is to model the observed state as a linear mixture of the true state and the maximally mixed state \cite{lvovsky2001}. Although with this assumption one can model certain simple noise sources, such as equal dark counts arriving with equal rates at each detector, we address the specific sources of noise in our experimental paradigm more directly. To model dark counts, we assume that the production of photons by the laser source, and the arrival of dark counts at the detectors can be modeled using independent homogeneous Poisson process with (constant) rate parameters $\lambda_s$ for the source and $\lambda_d^\gamma$ for each detector. These rates are estimated a priori using a pilot experiment. From these assumptions one can derive the following likelihood function using the properties of the Poisson distribution:
\begin{equation}\label{likelihood_dark}
\mathbb{P}(\gamma|\rho, \alpha, \lambda_s, \lambda_d^1,\ldots,\lambda_d^\gamma) =
\frac{\mathrm{Tr}[M_{\alpha\gamma}\rho]\lambda_s + \lambda_d^\gamma}
{\lambda_s + \sum_\gamma \lambda_d^\gamma}
\end{equation}

To deal with channel efficiency, we assume that there is a fixed probability of a photon being ``lost'' from a channel; this probability is denoted $1-\eta_\gamma$ for each channel $\gamma$, and hence $\eta_\gamma$ may be interpreted as the ``channel efficiency''. These probabilities are also estimated in a preliminary experiment. Given these efficiencies the likelihood becomes:
\begin{equation}\label{likelihood_inefficient}
\mathbb{P}(\gamma|\rho, \alpha, \eta_1,\ldots,\eta_\gamma) =
\frac{\mathrm{Tr}[M_{\alpha\gamma}\rho]\eta_\gamma}
{\sum_\gamma \mathrm{Tr}[M_{\alpha\gamma}\rho]\eta_\gamma}
\end{equation}

Note that in both cases, both the numerator and denominator contain only linear terms in the additional parameters ($\lambda, \eta$). Therefore, one only requires estimates of the ratio of the dark count rates to the source rate $\lambda_d^\gamma/\lambda_s$, and, for single-qubit tomography, the ratio of the efficiencies of the two channels $\eta_1/\eta_2$. It is straightforward to show that this property also holds when one collects blocks of measurements in one configuration and computes the likelihood of all measurements in the block simultaneously.

\begin{figure}[h!]
\includegraphics[width=\columnwidth]{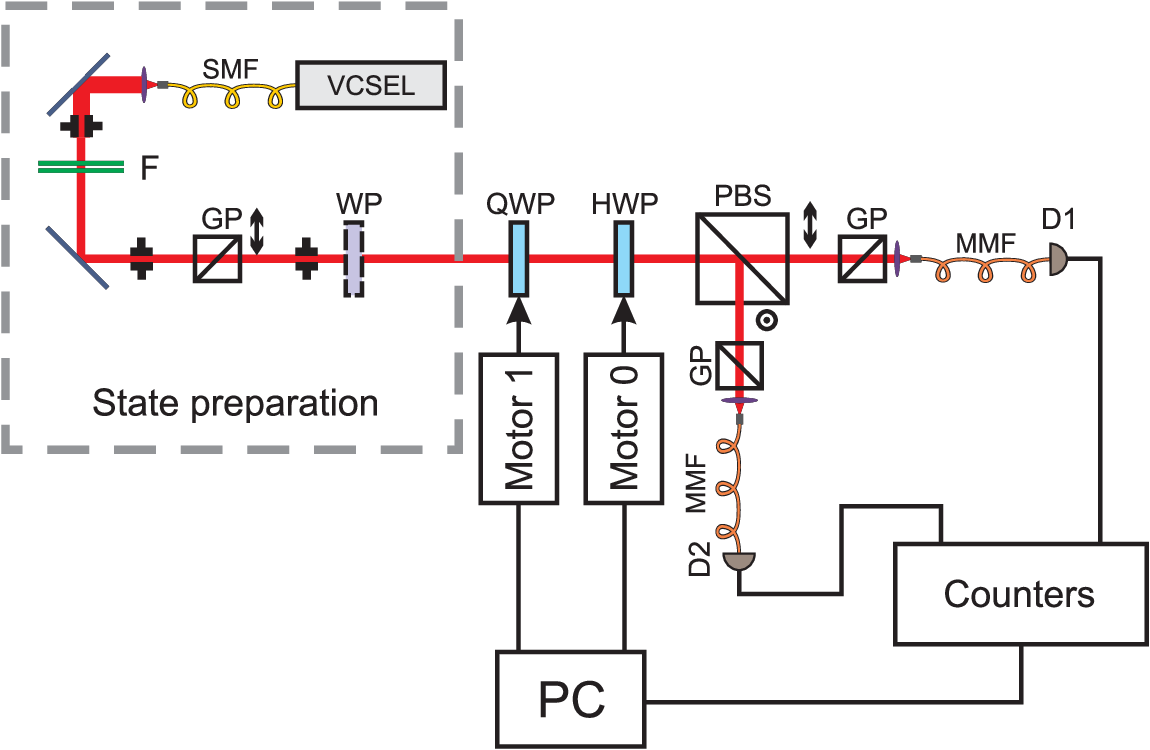}
{\caption{Experimental setup. An attenuated laser is used as a source, the polarization state is prepared by a set of waveplates, and analyzed by a sequence of a quarter- and half-wave plates, followed by a polarizing beam-splitter and two single-photon counters. Waveplates are rotated by electronically controlled step-motor drivers to allow for adaptivity.}\label{Setup}}
\end{figure}

\paragraph{Experiment.}
A sketch of our experimental setup is shown in Fig. \ref{Setup}. We use a CW 850 nm VCSEL diode laser coupled to a single-mode fiber as a light source. The radiation is attenuated to the single-photon level by a set of neutral density filters F and additionally spatially filtered with small iris apertures. The input polarization state is defined by a Glan-Taylor prism GP with high extinction ratio (more than 6000:1), the prism transmits horizontally polarized light, which may be transformed to some arbitrary state with a custom quartz plate WP.

The measurement scheme consists of zero order quarter-wave plate QWP and half-wave plate HWP. The plates are rotated by step-motor-driven stages, with minimal angular step of $0.1^\circ$. The zero position is controlled by a Hall sensor providing uncertainty of wave-plates zero of $0.2^\circ$. We clean up the polarization states in the output channels of a polarization beam-splitting cube (PBS) with two additional Glan-Taylor prisms to ensure high extinction ratio. Effectively that is equivalent to introducing some losses in the non-ideal PBS without altering the output polarization states. In each channel photons are coupled to multi-mode fibers and detected by single photon counting modules D1 and D2 (Perkin-Elmer). Electronic pulses from SPCM's are sent to home-made counters which may operate in two regimes~-- count for a fixed period of time or count until the specified number of counts is reached.

\begin{figure}[h!]
\includegraphics[width=\columnwidth]{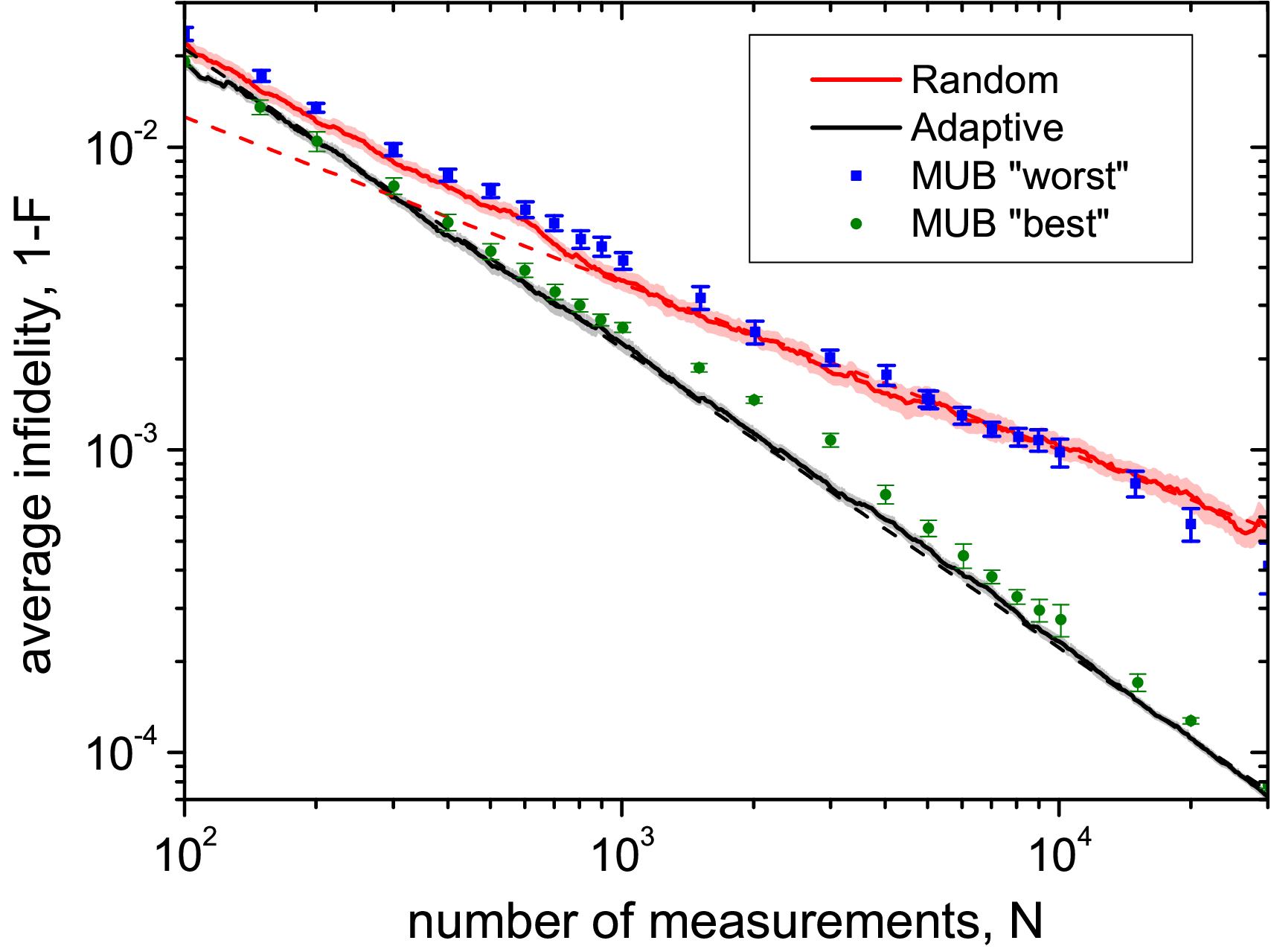}
{\caption{Experimental results: mean infidelity $1-\hat{F}(\rho,\hat{\rho})$ with current Bayesian mean estimate $\hat{\rho}$ for random measurements~-- red (upper) line, adaptive measurements~-- black (lower) line, measurements in ``worst'' MUBs~-- blue (dark grey) points and in ``best'' MUBs~-- green (light grey) points. Data points are averaged over 10 experimental runs, shaded areas and error bars show standard deviation of mean. Dashed straight lines are power law fits.}\label{Infidelity_to_estimate}}
\end{figure}

%%%%%%%%%%%%%%%%%%%%%
%Gleb: дополнено и перекомпоновано
To show the advantage of adaptive state estimation over non-adaptive protocols we performed a direct comparison in the Bayesian framework. In the adaptive estimation scheme we used two strategies: adaptation after every single measurement and block measurements and found no statistically significant differences. Fig. \ref{Infidelity_to_estimate} shows the dependence of mean infidelity $1-\hat{F}(\rho,\hat{\rho}) = \mathbb{E}_{p(\rho|\mathcal{D}_n)}\left[ 1-F(\rho,\hat{\rho}) \right]$ with current estimate $\hat{\rho}$ (for which we used the Bayesian posterior mean) on the number of measurements performed.
It is important to note that we intentionally \emph{did not average} over many realizations at each step of the algorithm, data points in Fig. \ref{Infidelity_to_estimate} are averaged over several \emph{full runs} of the experiment. The convergence rate behaves regularly from run to run. Fitting the data averaged over 6 realizations with power law of the form $1-\hat{F}\propto N^a$, we obtained $a=-0.700\pm0.005$ for random measurements, $a=-0.889\pm0.003$ for adaptive protocol, $a=-0.949\pm0.003$ for measurements in ``best'' MUBs, and $a=-0.680\pm0.003$ in ``worst'' MUBs. All non-adaptive protocols were shown to scale similarly in the limit of large $N$ in our simulations, except measurements in the eigenbasis of the state under estimation~\cite{Steinberg} (``best'' MUBs are a specific kind of them). So scaling for random and ``worst'' MUB protocols is typical for all static protocols.

Knowledge of the true state is essential to construct the ``best'' MUBs. But in a real world application of tomography the ``true'' state is unknown, and the Bayesian estimate given above is the only figure of merit at hand. In our experiment we have averaged over 6 runs of the adaptive protocol and used the result as an estimate of the ``true'' state's density matrix, the ``best'' MUBs were aligned with its eigenvector with nearly unit eigenvalue.

\begin{figure}[h!]
\includegraphics[width=\columnwidth]{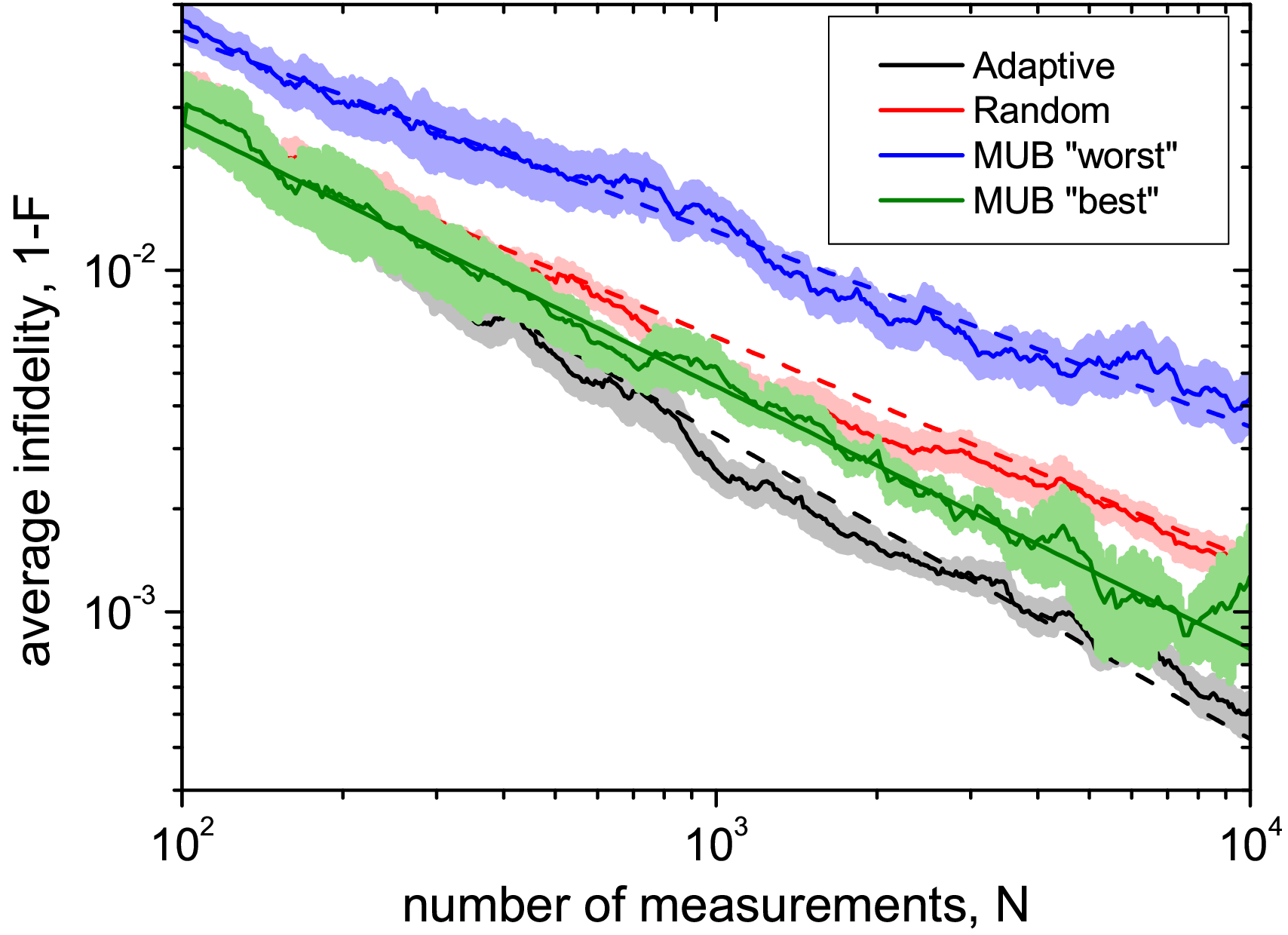}
{\caption{Experimental results: mean infidelity $1-\hat{F}(\rho,\bar{\rho})$ with ``true'' state $\bar{\rho}$ for random measurements~-- red (upper medium) line, adaptive measurements~-- black (the lowest) line, measurements in ``worst'' MUBs~-- blue (topmost) line and in ``best'' MUBs~-- green (lower medium) line. Data points are averaged over 10 experimental runs, shaded areas show standard deviation of mean. Dashed straight lines indicate the power law fits.}\label{Infidelity_to_true}}
\end{figure}

We may now analyze the protocol performance using the mean infidelity with the ``true'' state $\bar\rho$: $1-\hat{F}(\rho,\bar{\rho}) = \mathbb{E}_{p(\rho|\mathcal{D}_n)}\left[ 1-F(\rho,\bar{\rho}) \right]$. Here for determining the ``true'' state another technique was applied: we performed a very large number of Stokes parameters measurements. The scaling with $N$ of mean infidelity with state, estimated this way, is depicted in Fig. \ref{Infidelity_to_true}. Power law fits give for random strategy $a=-0.502\pm0.001$, for adaptive protocol $a=-0.902\pm0.008$, for measurements in aligned MUBs $a=-0.771\pm0.003$ and in ``worst'' MUBs $a=-0.589\pm0.007$. Error bars here are from the fit of an average curve, and within errors experimentally obtained scaling laws agree with simulations.

\begin{figure}[h!]
\includegraphics[width=\columnwidth]{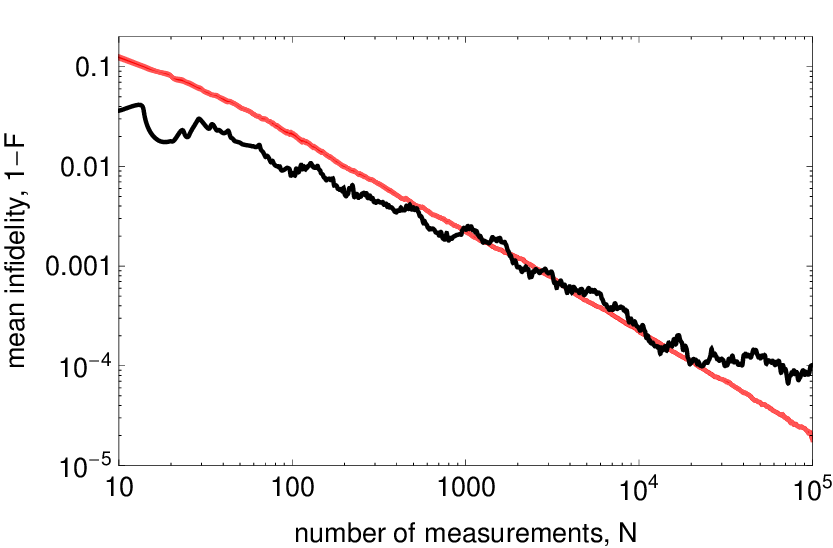}
{\caption{Comparison of mean distribution size $1-\hat F(\rho, \hat\rho)$ (red) and actual spread of the results $1-F_s$ (black) using 10 adaptive tomography runs.}\label{Spread}}
\end{figure}

%Gleb: Новый раздел
\paragraph{Systematic error assessment.}
Mean infidelity with current Bayesian mean estimate $1-\hat F(\rho, \hat \rho)$ (distribution size) shows the statistical uncertainty of the tomography result. It is a natural error estimate in a Bayesian framework. On the other hand, having performed several runs of tomography we can compute an actual spread of the results. Let $\hat \rho_k(N)$ be the Bayesian mean estimate in the $k$-th experimental run. Then the actual spread is determined as follows: $1-F_s(N)=1 - \frac1K \sum_{k=1}^K F(\hat \rho_k(N), \sigma(N))$, where $\sigma(N)=\frac1K \sum_{k=1}^K \hat \rho_k(N)$ is the mean of estimates on $N$-th step over $K$ runs. The results of this calculation for $K=10$ adaptive tomography runs are presented in Fig.~\ref{Spread}. For $N \lesssim 2\times10^4$ the two quantities are close to each other, as expected, but for $N \gtrsim 2\times10^4$ the actual spread becomes larger then the distribution size and reaches a steady value of $1-F_{s,min}\approx10^{-4}$. Such behavior may be attributed to systematic experimental errors which were not taken into account in our consideration so far. 

\begin{figure}[h!]
\includegraphics[width=\columnwidth]{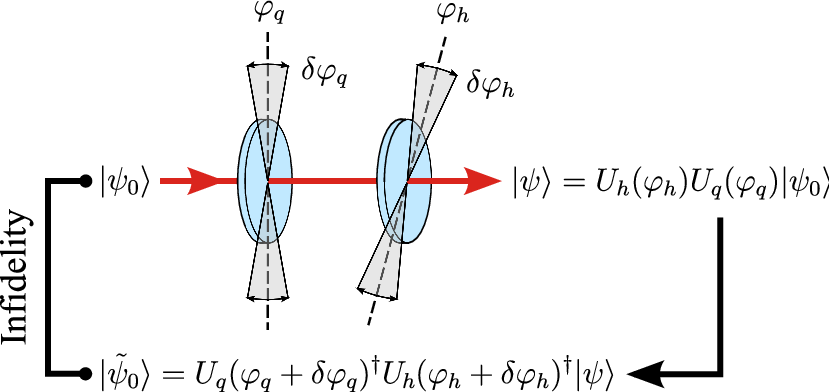}
{\caption{Diagram illustrating the estimation of systematic errors due to inaccuracies in waveplates rotation. True state $\ket{\psi_0}$ is transformed to $\ket{\psi}$ by the waveplates. The inverse image $\vert\tilde\psi_0\rangle$ of $\ket{\psi}$ is determined assuming that the waveplates have errors $\delta\varphi_{q,h}$ in angle position and infidelity with the true input state is calculated. }\label{RotationErrors}}
\end{figure}

One of the main sources of systematic errors is the inaccuracy in positioning of the wavepaltes. To estimate the value $1-F_{s,min}$ let us limit ourselves with pure states (almost pure ones were used in our experiment anyway). Let $\ket{\psi_0}$ be a true state and $\varphi_{q,h}$~-- the waveplates angles. Under the action of the waveplates the state is transformed to $\ket{\psi}=U_h(\varphi_h) U_q(\varphi_q) \ket{\psi_0}$, where $U_{q,h}$ are the unitaries, corresponding to transformations of the quarter- and half-wave plates respectively. Since the precise values of waveplates angles are unknown, the adaptive algorithm uses somewhat different values $\varphi_{q,h}+\delta \varphi_{q,h}$ -- the actual settings for the motors firmware. Thus the algorithm will not converge to a true state $\ket{\psi_0}$, but rather to $\vert\tilde\psi_0\rangle=U_q^\dagger(\varphi_q + \delta\varphi_q) U_h^\dagger(\varphi_h + \delta\varphi_h) \ket{\psi}$ which is the inverse image of $\ket{\psi}$ with inaccurate values of angles $\varphi_{q,h}+\delta \varphi_{q,h}$ in waveplates unitaries (see Fig.~\ref{RotationErrors}). Maximizing infidelity $1-F(\ket{\psi_0}, \vert\tilde\psi_0\rangle)$ between the true and the ``seeming'' states over angle inaccuracies in the interval $\vert \delta\varphi_{q,h} \vert \leqslant \delta \varphi$ determined by experimental imperfections and over all possible true states one achieves an upper bound on the infidelity $1-F_{s,min}$ due to systematic errors.

Numerical calculations following the described procedure give the value of infidelity $1-F=1.2 \times 10^{-4}$ for $\delta\varphi=0.2^{\circ}$~-- the experimentally determined inaccuracy of waveplates zero position. This result is in good agreement with the saturation value $1-F_{s,min}= 10^{-4}$ obtained from Fig.~\ref{Spread}.

%%%%%%%%%%%%%%%%%%%%%

\paragraph{Conclusion.}
Our experimental results clearly demonstrate the advantages of adaptive strategies in quantum state tomography. We have adapted Bayesian methods of state estimation, because Bayesian methods maintain confidence levels, and error bars with their estimates, they are a very natural tool for the task of adaptive experiment design. Besides the aforementioned favorable properties, the Bayesian approach is convenient from a purely practical point of view. It does not require any additional precomputation, and since posterior updates may be easily carried out after a single detection event, we expect that this approach will be particularly useful in the case of extremely weak signals. The $N^{-1}$ scaling of infidelity in the adaptive case is the theoretical limit for any tomographic protocol, and further improvement may only affect pre-factors in this power law. Simulation results show that our strategy of choosing adaptively between general measurements outperforms any non-adaptive protocol, and although only results for completely random measurements and fixed bases are provided here, experimental work showing worse performance of more sophisticated non-adaptive strategies is underway. Finally, let us note that using an attenuated laser source is absolutely equivalent to a true single-photon source for the purposes of this particular single qubit experiment. Generalization of the developed adaptive protocol for two-qubit polarization states, and higher-dimensional systems (like spatial modes of the biphoton field) will be reported elsewhere.

After this paper was completed we have become aware of a highly relevant work \cite{Steinberg} taking a different approach to adaptive state estimation and achieving similar performance.

This work was supported in part by Federal Program of the Russian Ministry of Education and Science (grant 8393) and state contract 911.519.11.4009, European Union Seventh Framework Programme under grant agreement n0 308803 (project BRISQ2), ERA.Net RUS Project NANOQUINT, grant for the NATO project EAP.SFPP 984397 "Secure Communication Using Quantum Information Systems" and by RFBR grants 12-02-31041 and 12-02-31792. S.~S.~Straupe is grateful to the "Dynasty" foundation for financial support. N. Houlsby is grateful for support from The Google European Doctoral Fellowship programme.

\end{document}